

\input phyzzx
\hsize=417pt 
\sequentialequations
\Pubnum={ EDO-EP-1 }
\date={ May 1994 }
\titlepage
\vskip 32pt
\title{ Black Hole Thermodynamics from String Theory }
\author{ Ichiro Oda }
\vskip 16pt
\address{ Edogawa University,                                 \nextline
          474 Komaki, Nagareyama City,                        \nextline
          Chiba 270-01, JAPAN                                 \nextline
            }
%
%
%
\abstract{
In this note we consider a stringy description of black hole
horizon. We  start with a nonlinear sigma model defined
on a two dimensional Euclidean surface  with background
Rindler metric. By solving the field equations, we show
that to the leading order the Bekenstein-Hawking formula
of black hole entropy can be produced.
We also point out a relation between
the present formalism and the 'tHooft formalism.
}

\endpage
%
%
%

\def\sp(#1){ \noalign{\vskip #1pt} }

%
%
\REF\HawkingI{
                S.W. Hawking,
		Comm.Math.Phys. {\bf 43}, 199 (1975)
             }
\REF\Pres{
                J. Preskill,
		Physica Scripta  {\bf T36}, 258 (1991)
		}
\REF\Bomb{
                L. Bombellli, R.K. Koul, J. Lee
                and R.D.Sorkin,
                Phys.Rev. {\bf D34}, 373 (1986)
		}
\REF\tHooftI{
                G. 'tHooft,
                Nucl.Phys. {\bf B256}, 727 (1985)
		             }
\REF\Sred{
                M. Srednicki,
	        Phys.Rev.Lett. {\bf 71}, 666 (1993)
             }
\REF\SussI{
                L. Susskind and J. Uglum,
	        Stanford preprint SU-ITP-94-1,
                hep-th/9401076
             }
\REF\CallanI{
                C. Callan and F. Wilczek,
	        IAS preprint IAS-HEP-93/87,
                hep-th/9401072
             }
\REF\Kabat{
                D. Kabat and M.J. Strassler,
	        Rutgers preprint RU-94-10,
                hep-th/9401125
             }
\REF\Jacob{
                T. Jacobson,
	        Maryland preprint, gr-qc-9404039
             }
\REF\Beken{
                J.D. Bekenstein,
                Nuovo Cim. Lett. {\bf 4}, 737 (1972);
                Phys.Rev. {\bf D7}, 2333 (1973);
                ibid. {\bf D9}, 3292 (1974);
                Physics Today {\bf 33}, no.1, 24 (1980)
	     }
\REF\tHooftII{
                G. 'tHooft,
	        Nucl.Phys. {\bf B335}, 138 (1990) ;
	        Physica Scripta {\bf T15}, 143 (1987) ;
	        ibid. {\bf T36}, 247 (1991);
                Utrecht preprint THU-94/02,
                gr-qc/9402037
             }
\REF\SussII{
                L. Susskind, L. Thorlacius and J. Uglum,
                Phys.Rev. {\bf D48}, 3743 (1993);
                L. Susskind and L. Thorlacius, Phys.Rev.
                {\bf D49}, 966 (1994)
             }
\REF\Unruh{
                W. G. Unruh,
                Phys.Rev. {\bf D14}, 870 (1976)
             }
\REF\Gibb{
                G. Gibbons and S.W. Hawking,
                Phys.Rev. {\bf D15}, 2752 (1977);
                S.W. Hawking, From General Relativity:
                An Einstein Centenary Survey,
                Cambridge Univ. Press 1979
             }
\REF\Oda{
                I. Oda,
                Int.J.Mod.Phys. {\bf D1}, 355 (1992)
             }
\REF\Magg{
                M. Maggiore,
                preprint IFUP-TH 22/94,
                hep-th/9404172
             }

%
%
\topskip 30pt
\par
To construct a theory of quantum gravity in four dimensions
is one of the most difficult and challenging subjects left
in the modern theoretical physics since we have so far
neither useful informations from experiments nor consistent
quantum field theory. Under such a circumstance, it seems to
be an orthodox attitude to attack concrete problems with
logical conflicts and then learn the fundamental principle
from which in order to construct a full-fledged theory.
In the case of quantum gravity, as one of such unsolved
problems, we have quantum black holes \PRrefmark{\HawkingI}.
In particular, it is widely known that there are at least
three problems which remain to be clarified in quantum black
hole, those are, the endpoint of Hawking radiation,
the information loss paradox and the statistical origin of black
hole entropy \PRrefmark{\Pres}.

Recently, there have been some progresses on the last problem
\PRrefmark{\Bomb-\Jacob}. Among them, the authers of
Ref.[\SussI] made an interesting observation that
superstring theory might  play an important role in deriving
the Bekenstein-Hawking formula of the black hole entropy
\PRrefmark{\HawkingI, \Beken}.

On the other hand, in previous works \PRrefmark{\tHooftII},
'tHooft has stressed that black holes are as fundamental as
strings, so that the two pictures are really complementary.
In fact, he has demonstrated that by properly taking account
of a leading gravitational back-reaction of the black hole
horizon, the gravitational shock wave, from hard particles,
his S matrix which describes the dynamical properties of
a black hole can be recast in the form of functional integral
over the Nambu-Goto string action. Although his formalism
has some weaknesses, it is extremely interesting from the
physical viewpoint since quantum incoherence never be lost
and all information of particles entering into a black
hole is transmitted to outgoing particles owing to the
Hawking radiation through the quantum fluctuations of
the black hole horizon.
As it is expected that superstring has many degrees of
freedom and hairs associated with its many excited states,
the 'tHooft formalism might also give us a clue to
understanding of a huge entropy \PRrefmark{\Beken}
and quantum hairs \PRrefmark{\Pres} of a black hole.

In this note, we shall simply assume that the dynamics
of the event horizon
of a black hole can be described by the world sheet
swept by a string in the Schwarzschild background,
and then would like to discuss what physical
consequences can be derived from this assumption.
However, the Schwarzschild metric is rather complicated,
so that we shall confine ourselves to the case
of the Rindler spacetime.
The case of the Schwarzschild metric will be
reported in a separate paper. We will see that
a nonlinear sigma action leads to the well-known
Bekenstein-Hawking formula of black hole entropy,
$S = {1 \over 4G} A_H$ \PRrefmark{\HawkingI, \Beken},
within the lowest order of approximation.
Moreover, one obtains a covariant operator algebra on the horizon
which is a natural generalization to the 'tHooft one
\PRrefmark{\tHooftII}. Thus our stringy approach to black hole
physics might be fruitful in both black hole thermodynamics
and 'tHooft formalism.

In relation to our stringy approach, some people might wonder
that the black hole horizon never be composed of a
string since a free falling observer crossing the horizon
encounters nothing unusual. However, this apparent contradiction
would be overcome by the principle of "black hole complementarity"
which has recently been advocated in Ref.[\SussII].
This principle says that the reference frame of an asymptotic
observer and that of a free falling observer approaching
the horizon of a black hole are very different and
the above-mentioned contradiction can be traced to unsubstantiated
assumptions about physics at or beyond the Planck scale.

As an effective action describing the dynamical properties
of the black hole horizon, we start with a classical action
of a nonlinear sigma model which is given by
\par
$$ \eqalign{ \sp(2.0)
S_E  & =   -{T \over 2} \int d^2 \sigma \sqrt{h}
h^{\alpha\beta} \partial_{\alpha} X^{\mu} \partial_{\beta}
X^{\nu} g_{\mu\nu} ( X ),
\cr
\sp(3.0)} \eqno(1)$$
where $T$ is a string tension having dimensions of mass squared.
$h_{\alpha\beta} ( \tau, \sigma )$ denotes the two dimensional
world-sheet metric which has a Euclidean signature, and $h \ = \
det \ h_{\alpha\beta}$. $X^{\mu} ( \tau, \sigma )$ maps the
string into a four dimensional spacetime, and then $g_{\mu\nu}
( X )$ can be identified as the background spacetime metric
 in which the string
is propagating. Note that $\alpha, \beta$ takes values 0, 1
and $\mu, \nu$ does values 0, 1, 2, 3.

The classical field equations give us that
$$ \eqalign{ \sp(2.0)
0 &= T_{\alpha\beta} = -{2 \over T} {1 \over \sqrt{h}}
{\delta S_E \over \delta h^{\alpha\beta}},
\cr
  &= \partial_{\alpha} X^{\mu} \partial_{\beta} X^{\nu}
g_{\mu\nu} ( X ) - {1 \over 2} h_{\alpha\beta} h^{\rho\sigma}
\partial_{\rho} X^{\mu} \partial_{\sigma} X^{\nu}
g_{\mu\nu} ( X ),
\cr
\sp(3.0)} \eqno(2) $$
$$ \eqalign{ \sp(2.0)
0 &= \partial_{\alpha} ( \sqrt{h} h^{\alpha\beta} g_{\mu\nu}
\partial_{\beta} X^{\nu} ) - {1 \over 2} \sqrt{h}
h^{\alpha\beta} \partial_{\alpha} X^{\rho} \partial_{\beta}
X^{\sigma} \partial_{\mu} g_{\rho\sigma}.
\cr
\sp(3.0)} \eqno(3) $$
In this note we consider the case that the background spacetime metric
$g_{\mu\nu} ( X )$ takes a form of the Euclidean Rindler metric
$$ \eqalign{ \sp(2.0)
ds^2 = g_{\mu\nu} dX^{\mu} dX^{\nu}
= + g^2 z^2 dt^2 + dx^2 + dy^2 + dz^2,
\cr
\sp(3.0)} \eqno(4) $$
where $g$ is given by $g \ = \ {1 \over 4M}$. This Rindler
metric can be obtained in the large mass limit from the
Schwarzschild black hole metric \PRrefmark{\Unruh}.
Here it is important to notice that we have performed the Wick
rotation with respect to the time component since now we would
like to discuss the thermodynamic properties of the Rindler
spacetime when we assume that the dynamics of the event
horizon is controlled by a Euclidean string.

Now one can easily solve Eq.(2)  as follows
$$ \eqalign{ \sp(2.0)
h_{\alpha\beta} = G(\tau,\sigma)\partial_{\alpha}
X^{\mu} \partial_{\beta} X^{\nu} g_{\mu\nu} (X),
\cr
\sp(3.0)} \eqno(5) $$
where $G(\tau,\sigma)$ denotes the Liouville mode.
Next we fix the gauge symmetries which are the two dimensional
diffeomorphisms and the Weyl rescaling by
$$ \eqalign{ \sp(2.0)
x(\tau,\sigma) = \tau, \  y(\tau,\sigma) = \sigma, \
G(\tau,\sigma) = 1.
\cr
\sp(3.0)} \eqno(6) $$
At this stage, let us impose an "axial" symmetry
$$ \eqalign{ \sp(2.0)
r(\tau,\sigma) = r(\tau), \ t(\tau,\sigma) = t(\tau).
\cr
\sp(3.0)} \eqno(7) $$
{}From Eq.s (5), (6) and (7),
the world sheet metric $h_{\alpha\beta}$ takes the form
$$ \eqalign{ \sp(2.0)
h_{\alpha\beta} = \left(\matrix{g^2 z^2 \dot t^2
+ \dot z^2  + 1  & 0 \cr
0 & 1  \cr} \right ),
\cr
\sp(3.0)} \eqno(8) $$
where the dot denotes a derivative with respect to $\tau$.
And the remaining field equations (3) become
$$ \eqalign{ \sp(2.0)
\partial_{\tau}
( \ { { z^2 \dot t } \over \sqrt{h}} \ )
 = 0,
\cr
\sp(3.0)} \eqno(9) $$
$$ \eqalign{ \sp(2.0)
\partial_{\tau} h = 0,
\cr
\sp(3.0)} \eqno(10) $$
$$ \eqalign{ \sp(2.0)
\partial_{\tau}
( \   {\dot z \over \sqrt{h}} \  )
- { 1 \over \sqrt{h}} g^2 z {\dot t}^2 = 0,
\cr
\sp(3.0)} \eqno(11) $$
where
$$ \eqalign{ \sp(2.0)
h = g^2 z^2 \dot t^2
+ \dot z^2  + 1.
\cr
\sp(3.0)} \eqno(12) $$

Now it is straightforward to solve the above field equations.
We have two kinds of solutions. One solution is a trivial
one given by
$$ \eqalign{ \sp(2.0)
z = \dot z = \ddot z = 0 , \ t(\tau) = arbitrary.
\cr
\sp(3.0)} \eqno(13) $$
which corresponds to a world-sheet surface of
the Euclidean string  just lying on the black hole horizon.
The next solution is the solution of "world sheet instanton"
described by
$$ \eqalign{ \sp(2.0)
z(\tau) =  \sqrt{ c_2 (\tau - \tau_0)^2 +
{g^2 c_1^2 \over c_2 } },
\cr
t - t_0 ={ 1 \over g} {\tan}^{-1} {c_2 \over {g c_1}}
(\tau - \tau_0),
\cr
\sp(3.0)} \eqno(14) $$
where $c_1 , c_2 , \tau_0$ , and $t_0$ are the integration
constants, in other words, "the moduli parameters".
To understand the physical meaning of this solution
more vividly, it is convenient to rewrite $z$ in terms
of the time coordinate variable $t$.
{}From Eq.(14), we obtain
$$ \eqalign{ \sp(2.0)
z(t_E) \ = \ { g c_1 \over \sqrt{c_2}}
{ 1 \over \cos{ g (t_E - t_{E0})}},
\cr
\sp(3.0)} \eqno(15) $$
where we added the suffix $E$ on $t$ in order to
indicate the Euclidean time clearly.
Furthermore after Wick-rerotating the time coordinate, we
have in the real Lorentzian time $t_L$
$$ \eqalign{ \sp(2.0)
z(t_L) \ = \ { g c_1 \over \sqrt{c_2}}
{ 1 \over \cosh{ g (t_L - t_{L0})}},
\cr
\sp(3.0)} \eqno(16) $$
The explicit form of the solution (16)
shows that this solution has a physical behavior
of approaching the event horizon $z = 0$
asymptotically in the Lorentzian time
coordinate $t_L$. Incidentally, note that
the solution (15) has a periodicity with
respect to the Euclidean time component, $\beta =
{ 2\pi \over g} $  whose inverse gives us nothing but
the Hawking temperature $T_H = {1 \over \beta}
= {g \over 2\pi} = {1 \over 8\pi M}$
of the Rindler spacetime \PRrefmark{\Unruh}.

Next we would like to evaluate the black hole
entropy to the leading order of approximation
within the present formalism by a method developed
by Gibbons and Hawking \PRrefmark{\Gibb}.
Before caluculation, let us
present a brief review of their method. The free
energy  $F$  of a black hole in equilibrium with
a radiation bath can be computed in terms of the
Euclidean path integral
$$ \eqalign{ \sp(2.0)
{\rm e}^{-\beta F} = \int_{\beta\hbar} {\it DX}
{\rm e}^{- {S_E \over \hbar}},
\cr
\sp(3.0)} \eqno(17) $$
where $S_E$ denotes the Euclidean action, and the
path integral is performed under the boundary condition
of being periodic in the Euclidean time with period
$\beta\hbar$. Then the black hole thermodynamics can
be recovered in the limit $\hbar \rightarrow 0$ by
expanding $S_E$ around its saddle point. Thus evaluating
the free energy $\beta$ to the leading term equals to
substituting a classical solution into the Euclidean
action.

In the model just considered, it is easy
to calculate the free energy. To do so we shall
consider the solution (14) since this solution
gives us the thermal temperature whose situation
should be contrasted to the case of the
other solution (13). The result is
$$ \eqalign{ \sp(2.0)
F =  - {1 \over \beta} \sqrt{c_2 + 1} \ T \ A_H,
\cr
\sp(3.0)} \eqno(18) $$
where $A_H = \int dx dy$ which corresponds to the area
of the black hole horizon if we consider the Schwarzschild
black hole.
By the formula which gives us the entropy
$$ \eqalign{ \sp(2.0)
S = {\beta}^2 \ {\partial F \over \partial \beta},
\cr
\sp(3.0)} \eqno(19) $$
one can show that the entropy is given by
$$ \eqalign{ \sp(2.0)
S = \sqrt{c_2 + 1} \ T \ A_H.
\cr
\sp(3.0)} \eqno(20) $$
Note that the black hole entropy
is proportional to the horizon area. Moreover,
by selecting the string tension
$$ \eqalign{ \sp(2.0)
T = {1 \over 4 \sqrt{c_2 + 1} G },
\cr
\sp(3.0)} \eqno(21) $$
we arrive at the famous Bekenstein-Hawking entropy
formula \PRrefmark{\HawkingI, \Beken}
$$ \eqalign{ \sp(2.0)
S = {1 \over 4G} \ A_H.
\cr
\sp(3.0)} \eqno(22) $$

Next we shall consider the relation between our model
and 'tHooft one \PRrefmark{\tHooftII, \Oda}.
According to 'tHooft, some quantum fluctuations of
the event horizon can be induced by hard particles
having a large amount of momenta. Thus let us introduce
"vertex operator" in the original action (1)
$$ \eqalign{ \sp(2.0)
S_E   =  &-{T \over 2} \int d^2 \sigma \sqrt{h}
h^{\alpha\beta} \partial_{\alpha} X^{\mu} \partial_{\beta}
X^{\nu} g_{\mu\nu} ( X )
\cr
&+  \int d^2 \sigma \sqrt{h} P^{\mu}X^{\nu}
g_{\mu\nu}(X),
\cr
\sp(3.0)} \eqno(23)$$
where it is now supposed that the target spacetime
has a Lorentzian signature. Taking a variation with
respect to $X^{\mu}$ gives us the equation of  $X^{\mu}$
as follows
$$ \eqalign{ \sp(2.0)
0 = T {1 \over \sqrt{h}} \partial_{\alpha}
( \sqrt{h} h^{\alpha\beta} \partial_{\beta} X_{\mu} )
  + P_{\mu} ,
\cr
\sp(3.0)} \eqno(24) $$
where we have replaced $g_{\mu\nu}$ with the flat
metric $\eta_{\mu\nu}$. This approximation would become
good when the black hole mass is large compared to the
Planck mass. Moreover, we have fixed the world sheet
metric $h_{\alpha\beta}(\tau,\sigma)$ to be the metric
on $S^2$. Therefore we obtain
$$ \eqalign{ \sp(2.0)
T \Delta_{tr}X^{\mu} + P_{\mu} = 0,
\cr
\sp(3.0)} \eqno(25) $$
where
$$ \eqalign{ \sp(2.0)
\Delta_{tr} = {1 \over \sqrt{h}} \partial_{\alpha}
(\sqrt{h} h^{\alpha\beta} \partial_{\beta}).
\cr
\sp(3.0)} \eqno(26) $$

Now we assume the following commutation relations
which are motivated by the 'tHooft work
$$ \eqalign{ \sp(2.0)
\Bigl [ X^{\mu}(\sigma) , P^{\nu}({\sigma}') \Bigr ]
= i {\eta}^{\mu\nu} {\delta}^{(2)} (\sigma -{\sigma}'),
\cr
\sp(3.0)} \eqno(27) $$
in other words,
$$ \eqalign{ \sp(2.0)
P^{\mu}(\sigma) = -i{\delta \over {\delta X^{\mu}(\sigma)}}.
\cr
\sp(3.0)} \eqno(28) $$
{}From (25) and (27), we have
$$ \eqalign{ \sp(2.0)
\Bigl [ X^{\mu}(\sigma) , X^{\nu}({\sigma}') \Bigr ]
= {i \over T}{\eta}^{\mu\nu}  f(\sigma, {\sigma}'),
\cr
\sp(3.0)} \eqno(29) $$
where the Green function $f(\sigma, {\sigma}')$ is defined as
$$ \eqalign{ \sp(2.0)
-\Delta_{tr} f(\sigma, {\sigma}')
=  {\delta}^{(2)} (\sigma -{\sigma}').
\cr
\sp(3.0)} \eqno(30) $$
The commutation relations (29) are just the covariant
generalization to the relations which 'tHooft have
obtained in the context of shock wave geometry
\PRrefmark{\tHooftII}. Eq.(29) strongly suggests
that distances between adjacent points on the black
hole horizon in real world would be quantized with units
of the order Planck scale. Thus we have succeeded in
deriving the Lorentz covariant versions of the 'tHooft
commutation relations by assuming that the
dynamics of the horizon of
a black hole is controlled by a Euclidean string.

In conclusion, the main point of this article has been
 to demonstrate
that a black hole dynamics, in particular, the black
hole thermodynamics can be understood in terms of string
theory. We have seen that this is the case, at least in
a specific field theoretical model. In this model, we
have made an approximation that the Schwarzschild
black hole can be described by the Rindler
spacetime under the condition of  the large
black hole mass.

It is interesting to compare our derivation of the
black hole entropy with that of Ref.[\SussI]
and [\Jacob]. There is a clear difference
in the calculation method. The authors in
Ref.[\SussI] and [\Jacob]
evaluated the black hole entropy by first inducing
the Einstein-Hilbert action with the surface term
and the possible higher derivative terms from
string theory or pregeometry, and then considering
the geometries containing conical singularities.
On the other hand, we have directly evaluated
the black hole entropy by using the string action
without inducing the gravitational action from it.

\vskip 1cm

{\bf Notes added}

During the preparation of this article, we noticed that
there is a recent work where the black hole is
described by the membrane theory \PRrefmark{\Magg}.

\vskip 1cm

\noindent
{\bf Acknowledgement}

We are grateful to K. Akama, N. Kawamoto, A. Sugamoto and
Y. Watabiki for valuable discussions.

\refout
\vfill
%
\bye